# The interface states in gate-all-around transistors (GAAFETs)

Yue-Yang Liu[1,*,+], Haoran Lu[2,+], Zirui Wang[2], Hui-Xiong Deng[1], Lang Zeng[3], Zhongming Wei[1,*], Jun-Wei Luo[1,*], Runsheng Wang[2,*]

1 State Key Laboratory of Superlattices and Microstructures, Institute of Semiconductors, Chinese Academy of Sciences, Beijing 100083, China
2 Institute of Microelectronics, Peking University, Beijing 100871, China
3 Fert Beijing Institute, MIIT Key Laboratory of Spintronics, School of Integrated Circuit Science and Engineering, Beihang University, Beijing 100191, China

* Correspondence authors:
  E-mail addresses: yueyangliu@semi.ac.cn, zmwei@semi.ac.cn, jwluo@semi.ac.cn, r.wang@pku.edu.cn
+ These authors contribute equally to this work.

**Abstract**—The atomic-level structural detail and the quantum effects are becoming crucial to device performance as the emerging advanced transistors, representatively GAAFETs, are scaling down towards sub-3nm nodes. However, a multiscale simulation framework based on atomistic models and *ab initio* quantum simulation is still absent. Here, we propose such a simulation framework by fulfilling three challenging tasks, i.e., building atomistic all-around interfaces between semiconductor and amorphous gate-oxide, conducting large-scale first-principles calculations on the interface models containing up to 2796 atoms, and finally bridging the state-of-the-art atomic level calculation to commercial TCAD. With this framework, two unnoticed origins of interface states are demonstrated, and their tunability by changing channel size, orientation and geometry is confirmed. The quantitative study of interface states and their effects on device performance explains why the nanosheet channel is preferred in industry. We believe such a bottom-up framework is necessary and promising for the accurate simulation of emerging advanced transistors.

# 1 Introduction

Gate-all-around MOSFETs (GAAFETs) have been recognized as the promising successor of FinFETs in the coming sub-3nm technology node. Superiorities of GAAFET such as better short channel control, higher DC performance, less complex patterning strategy, and better design flexibility have been demonstrated successively in recent years [1–7]. Nevertheless, superior does not mean perfect. Significant performance degradations, including the rise of resistance, increase of subthreshold swing, aggravation of drain induced barrier lowering, and worsening of hot carrier reliability, have been observed when the nanowire channel is scaled to less than 4nm in diameter [8,9]. The issues were demonstrated more than a decade ago and were attributed to the aggravation of interface scatterings, but no follow-up microscopic investigation on the interface states, such as their intrinsic origins, energetic and spatial distributions, and tunability, is shown even up to now. The most related works have been several pioneering semi-empirical calculations on interface roughness scattering [10–12] and phonon scattering [13] [14]. At least two facts should be responsible for this situation. First, the GAAFETs with sub-4nm channel are too small for experimental techniques to characterize the interface precisely. On the other hand, the GAAFETs are too large and complex for atomic-level *ab initio* approaches to calculate accurately. Such a dilemma calls for the development of advanced characterizing or modeling/simulation techniques.

The challenge of simulating GAAFET from atomic level to device level is threefold. First, an atomistic GAAFET model with sufficient size (comparable to the state-of-art process) and high-quality all-around interface between Si and amorphous $SiO_2$ is difficult to build. To date, most of the atomistic simulation on semiconductor-oxide interfaces are based on planar interface models, e.g. $Si/SiO_2$, $Si/SiO_2/HfO_2$ and $MoS2/SiO_2$ [15–21]. Second, even if the atomistic model of GAAFET is obtained somehow, the number of atoms (N) in the model could be too large for first-principles calculation. The common commercial codes can only deal with systems with less than 1000 atoms, and the computational complexity scales up by $O(N^3)$. Finally, how to bridge the atomistic calculation with TCAD simulation is uncertain. The first-principles calculation can provide many physical quantities such as energy levels, wavefunctions, density of states (DOS) and so on. Which microscopic quantity can be used by the TCAD tool to generate macroscopic ones, such as I-V curves and subthreshold swings, is to be determined.

In this paper, we endeavor to propose strategies to deal with the above challenges one by one, and finally come up with a multiscale simulation framework for accurate simulation of GAAFETs. With this framework, the origin, property and tunability of interface states in $Si$-$SiO_2$ GAAFETs are revealed, and their effects on device performance are quantitatively presented. Furthermore, the feasibility of improving device performance by suppressing interface states is confirmed.

## 2 Modeling and Simulation Methods
### 2.1 High-quality Atomistic Modeling

A GAAFET is typically composed of a Si nanowire channel and an annular $SiO_2$ dielectric, which covers around the channel and is amorphous in structure. The Monte Carlo bond-switching (MC-BS) method [22–24] is adopted to build the GAAFET models with high-quality interface between Si and amorphous $SiO_2$, as is shown in Fig. 1(a)(b). The method requires an initial crystalline structure with correct bond information. Then, as illustrated in Fig. 1(c), the neighboring bonds will be randomly selected and broken at each step of the simulation, and then two new bonds are generated and followed by a relaxation of the whole system. Whether this bond-switch move will be accepted or not is determined by the energy change. Such a bond-switching process will be repeated for about 260,000 times with decreasing temperatures to obtain a fully disordered structure. The quality of the generated nanosheet model is shown in Fig. 1(d). It can be seen that the Si-O and O-O radial distribution function peaks at 1.62 Å and 2.60 Å, respectively, which is in good agreement with the experimentally reported 1.6 Å and 2.6 Å [25]. The O-Si-O angles peak at 108.5°, which corresponds well to the tetragonal bond angle. The coordination number of O atoms is exactly 2, indicating that there are no O dangling bonds. The coordination of Si atoms ranges from 0 to 4, depending on their location at bulk Si, at the Si-$SiO_2$ interface or at the $SiO_2$, respectively.

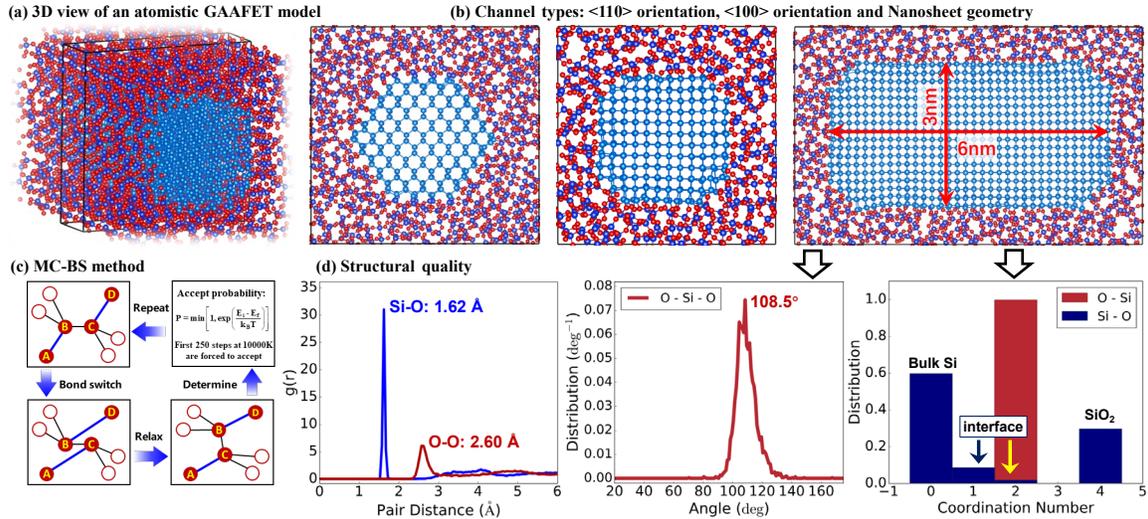

**Fig. 1** The building of atomistic GAAFET models. (a)-(b) Model configurations. (c) Construction method. (d) Quality check.

Note, the bond-switching simulation and DFT calculation will be more and more computationally-expensive with increasing number of atoms, thus it is impossible to build and calculate a nanowire that is as long as the experimentally grown ones. Eclectically, we study the models that are about 1nm-thick and periodic along the longitudinal direction. The final thickness of the [110] and [100] supercells are relaxed to be 11.65 Å and 10.94 Å, respectively. With such a compromise, the largest model contains 2796 atoms, which reaches the limitation of the DFT simulation capability. Through the above workflow, a series of GAAFET models with different diameters, shapes, and orientations have been constructed and studied.

## 2.2 Large-scale First-principles Calculations

The multiscale simulation contains two main parts, namely the atomic level first principles calculation for electronic properties, and the device level TCAD simulation for device performances. The connection of the two parts is realized by unifying the structural parameters and converting the output of atomic level calculation to the input of device simulation. The workflow is illustrated in Fig. 2. To enable the first-principles calculation on large GAAFET models, the *PWmat* package [26,27] that is optimized on GPU architectures to accelerate DFT calculation is utilized. 88 GPU cards (NVIDIA GTX 1080Ti) in parallel is found to be necessary to simulate the largest model with 2796 atoms. To reduce the computational complexity, an optimized norm-conserving Vanderbilt pseudopotential with no semicore valence electrons is used [28], by which the cutoff energy can be reduced to 45 Ry. This pseudopotential produces a lattice constant of 5.46 Å for silicon, which is in good agreement with the experimental value of 5.43 Å. All the large models have been optimized with a force converge criteria of 0.02 eV/Å. The GGA-PBE functional is used throughout the calculations.

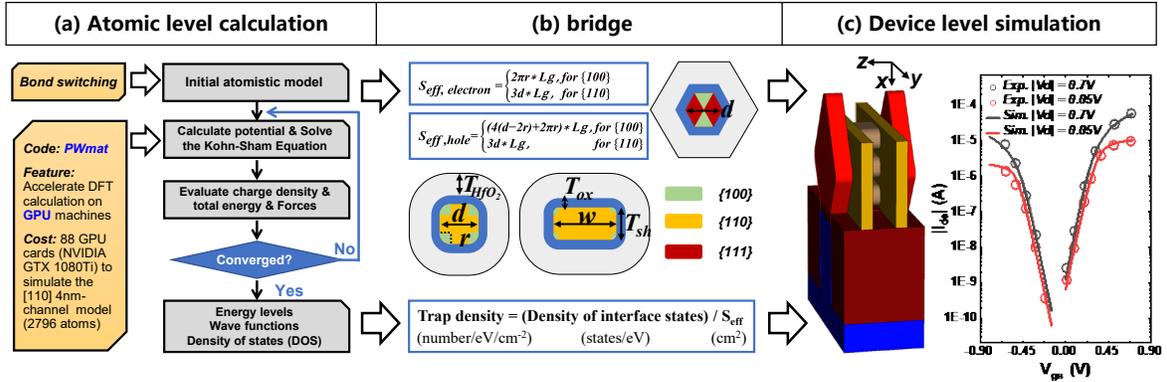

**Fig. 2** The work flow of the multiscale simulation framework. (a) The large-scale first-principles calculation. (b) The bridge between first-principles calculation and TCAD simulation. (c) The GAAFET model in TCAD simulation and the precise calibration.

## 2.3 TCAD simulation

The Sentaurus TCAD is used to investigate the effect of interface states on device performance. First, as is shown in **Fig. 2(b)**, the cross-section of the GAAFET channel is carefully designed according to the atomistic structures. For <100> oriented channels, the cross-sections are hexagons. For <110> orientations, channels are squares or rectangle with round corners. Second, the density of interface states (with the unit of states/eV) obtained from first-principles calculation is converted to trap density (with the unit of number/eV/cm$^{-2}$) by dividing the DOS by interface area ($S_{eff}$). This trap density is an input parameter of the TCAD simulation, and it will be summed over energy to calculate the total trapped charge. The detailed device parameters can be found in Table I. It is worth to mention that the physical models adopted in TCAD have been precisely calibrated based on the experimental results of manufactured 3D Stacked Nanosheet GAAFET in [3], as shown in **Fig. 2(c)**.

Table 1 Calibrated device parameters for TCAD simulation

| Geometric Parameters | Value | |
|---|---|---|
| Gate Length, $L_g$ (nm) | 12 | |
| S/D Epitaxial Growth Length, $L_{epi}$ (nm) | 9 | |
| Nanowire Diameters, $d$ (nm) | 1/2/3/4 | |
| Nanosheet Width, $w$ (nm) | 6 | |
| Nanosheet Thickness, $T_{sh}$ (nm) | 3 | |
| Gate Oxide Thickness, $T_{ox}$ (nm) | 0.9 | |
| **Doping Parameters** | **NMOS** | **PMOS** |
| Source Doping Concentration (cm$^{-3}$) | $2\times10^{19}$ | $2.5\times10^{19}$ |
| Drain Doping Concentration (cm$^{-3}$) | $2\times10^{19}$ | $2.5\times10^{19}$ |
| Channel Doping Concentration (cm$^{-3}$) | $2\times10^{15}$ | $1\times10^{15}$ |

### 3 Results and Discussion

3.1 The three origins of interface states.

**Fig. 3** shows the microscopic properties obtained by first-principles calculations. The total DOS (black line) is projected to the interfacial Si atoms to get the PDOS (red line), by which we can see how many states have been contributed by the interfacial atoms. These states are treated as interface states. **First**, it is found in Fig. 3(b) that the charge density (quadratic sum of the wave functions within an energy range) near conduction band minimum (CBM) localizes strongly at the four corners of the Si channel. This corresponds with the fact that the Si-Si bonds at the corners are longer than that inside bulk Si, as is shown in Fig. 3(c), indicating that the local strains are responsible for the corner states. Such tensile strain has been experimentally confirmed to exist in GAAFETs [29,30], and the band gap decrease induced by tensile strain has been reported in Si nanowires [31]. **In contrast,** the interface states near valence band maximum (VBM) distribute evenly along the whole circular interfaces, as is shown in Fig. 3(d), indicating a universal upward band bending at the interface. This is attributed to the formation of dipoles between Si and SiO$_2$, which stems from the electron migration from Si towards O due to their higher electronegativity. In fact, the GAAFET is a 3D potential well, as is shown in Fig. 3(e), in which the bands bend upwards near the interface. **Finally,** some very localized interface states can be seen when Si dangling bonds exist. Their energy levels locate at the band gap, which are consistent with the measurements in planar MOSFETs [32].

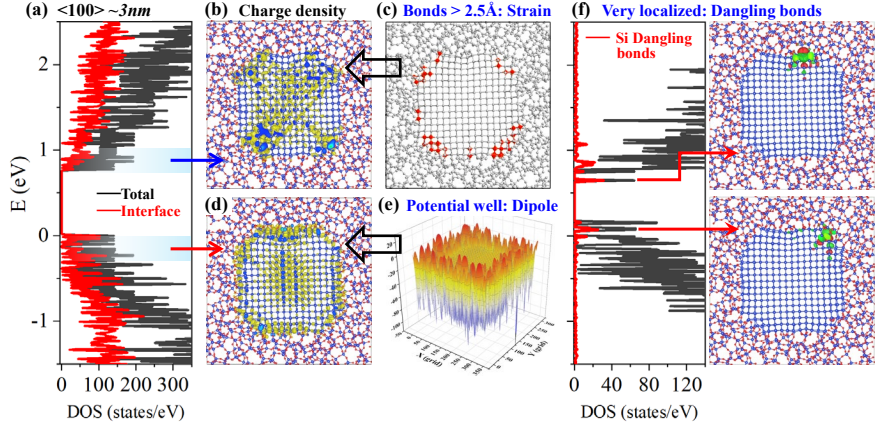

**Fig. 3** The microscopic quantities obtained from first-principles calculation. The spatial and energetic distribution of interface states indicate 3 microscopic origins: strains, dipoles and dangling bonds.

3.2 The tunability of interface states

The interface states are detrimental to the performance and reliability of GAAFETs, thus it is urgent to see whether the interface states can be reduced or tuned. For this purpose, we build more GAAFET models with different channel sizes, orientations and geometries. **Fig. 4(a)** shows the dependence of interface states on channel size. It can be seen that the ratio of interface states is very large for the models with 1-2 nm channel, indicating the challenge of keeping good performance in these small devices. Fortunately, the interface states decrease greatly, especially near the CBM, when the nanowire's diameter is increased to 4nm. These results are consistent with the experimental report that the GAAFETs begin to degrade when the channel decreases to 3nm or smaller [8]. **Fig. 4(b)** shows the change of channel orientation from <100> to <110>, this reduces the interface states near CBM further, but not effective to those near VBM. Excitingly, as is shown in **Fig. 4(c)**, the shift of nanowire channel to nanosheet one (with similar cross section area) significantly reduce the interface states around both CBM and VBM, indicating the great advantage of nanosheet configuration in boosting device performance. This should be one of the reasons that the nanosheet channel is preferred in industry.

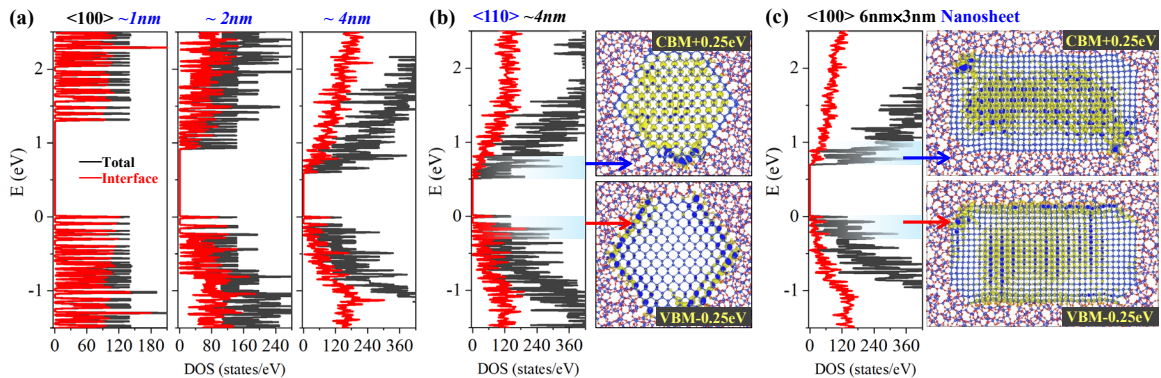

**Fig. 4** The effect of (a) channel size, (b) channel orientation, and (c) channel geometry on interface states.

## 3.3 Device performances

Based on the microscopic properties of interface states discussed above, a quantitative study of their effects on device performance is conducted. To distinguish the interface states with different origins, those induced by strains and dipoles are denoted as IF, while those by dangling bonds are denoted as DB. The interface trap density ($D_{it}$) of IF part is all obtained from our first-principles calculation, while that of the DB part is extracted from experiment [32]. **Fig. 5(a)** shows the different effects of IF and DB on current degradation of <100> NMOS. DB leads to shift of threshold voltage ($V_{th}$) while IF reduces current above threshold voltage. Different from NMOS, not only threshold voltage but also subthreshold swing (SS) of <100> PMOS is affected by DB in **Fig. 5(b)**, which can be explained as follows: When $V_{gs} = 0$, shallow hole traps below the fermi level are almost unoccupied. As $|V_{gs}|$ increases, more hole traps are occupied, leading to gradual current degradation, namely a deterioration of subthreshold swing and threshold voltage. For NMOS, however, electron traps with deeper energy level than hole traps are almost completely occupied when $V_{gs} = 0$, which manifests as a pure threshold voltage shift.

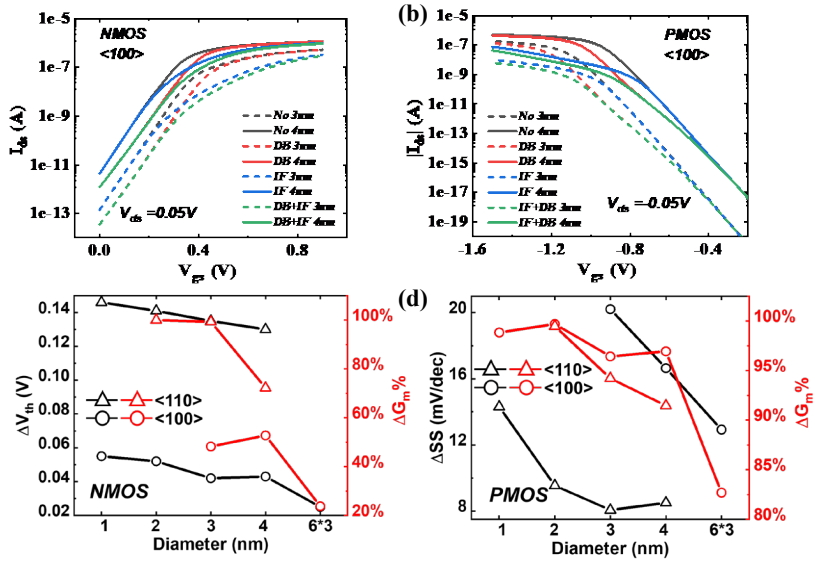

**Fig. 5** The device performances influenced by interface states. (a) The simulated $I_{ds}$ at $V_{ds}$=0.05V of NMOS with different traps combinations. DB (Red line) contributes to $V_{th}$ shift while IF (Blue line) contributes to current reduction above $V_{th}$. (b) The simulated $I_{ds}$ at $V_{ds}$=-0.05V of PMOS. (c) The increase of $G_m$ degradation and $\Delta V_{th}$ with diameter scaling. (d) The $\Delta$SS increases along with $G_m$ degradation in PMOS. The 6×3 nm nanosheet exhibits the best performance.

We use the relative change of transconductance ($\Delta G_m$) at the voltage where fresh GAAFETs have the maximum transconductance to characterize the effect of IF on current degradation above threshold voltage. IF electron traps are to be occupied after Fermi level enters the conduction band, resulting in a decline in drain current, namely a transconductance drop. As shown in **Fig. 5(c)**, the transconductance decreases with diameter scaling due to more IF portion. Bandgap widening [33] that comes with the smaller diameter leads to deeper DB distribution, so the shift of threshold voltage ($\Delta V_{th}$) of NMOS increases. Compared with <100> orientation, <110> NMOS show larger threshold voltage shift, which is consistent with more DB at {111} interface. As the diameter becomes smaller, the degradation of subthreshold voltage ($\Delta SS$) increases along with transconductance degradation in PMOS in **Fig. 5(d)**. This is because deeper hole traps of DB result in more $D_{it}$ distribution in subthreshold. It is worth pointing out that nanosheet exhibits best performance in all respects, in agreement with the experimental report [34].

## 4 Conclusions

In conclusion, we have proposed a bottom-up framework that composed of high-quality atomistic modeling, large-scale first-principles calculation (on 2796 atoms), and TCAD simulation, to study the interface states and performance of emerging GAAFETs. The interface states exhibit a complicated distribution in both real space and energy space, and show an obvious dependence on channel size, channel orientation, and channel shape. Nevertheless, they can be classified into three categories, i.e. the well-known dangling bond induced ones that can be removed by hydrogen passivation, the unnoticed strain-induced ones that can be reduced by engineering the channel geometry, and the unnoticed dipole-induced ones that can hardly be tuned. The overwhelming interface states indicate that it is very challenging to maintain the high performance of nanowire GAAFETs at sub-3nm scale, especially for p-type devices. However, impressively, the shift of nanowire channel to nanosheet channel (with similar cross-sectional area) can significantly suppress the effect of interface states, and bring forth good device performance. It is thus practical and advisable for the industry to keep focusing on nanosheet GAAFETs.

**Acknowledgement**

This work was supported by the National Natural Science Foundation of China (Grants No. 62174155, No. 61927901, No. 62125404 and No. 12004375)